\documentclass[12pt]{article}
\usepackage[spanish,activeacute]{babel}
\usepackage{latexsym}
\usepackage[dvips]{graphicx}
\usepackage[latin1]{inputenc}%% PARA PODER ESCRIBIR TILDES, Ñ, ö, ect.
\usepackage{verbatim}%% PARA HACER COMENTARIOS CON EL ENTORNO comment
\usepackage{fancyhdr}%% PARA DEFINIR ESTILOS DE PAGINA Y ENCABEZADOS
\usepackage{makeidx}%% PARA HACER EL INDICE DE MATERIAS
\usepackage{amssymb}
\usepackage{amsmath}
\usepackage{amsthm}
\usepackage{array}
\usepackage{pst-all}
\usepackage{pstcol}
\usepackage{pst-lens}
\usepackage{pifont}
\usepackage[centerlast,small,sc]{caption2}
\usepackage{subfigure}
\usepackage{epsf}
\definecolor{dorado}{cmyk}{0,0.1,0.84,0}
\definecolor{melon}{cmyk}{0,0.29,0.84,0}
\definecolor{naranja}{cmyk}{0,0.42,1,0}
\definecolor{durazno}{cmyk}{0,0.46,0.5,0}
\definecolor{salmon}{cmyk}{0,0.53,0.38,0}
\definecolor{fresa}{cmyk}{0,1,0.5,0}
\definecolor{ladrillo}{cmyk}{0,0.77,0.87,0}
\definecolor{violeta}{cmyk}{0.07,0.9,0,0.34}
\definecolor{purpura}{cmyk}{0.45,0.86,0,0}
\definecolor{turquesa}{cmyk}{0.85,0,0.2,0}
\definecolor{aguamarina}{cmyk}{0.85,0,0.33,0}
\definecolor{esmeralda}{cmyk}{0.91,0,0.88,0.12}
\definecolor{pino}{cmyk}{0.92,0,0.59,0.25}
\definecolor{oliva}{cmyk}{0.64,0,0.95,0.4}
\definecolor{canela}{cmyk}{0.14,0.42,0.56,0}
\definecolor{marron}{cmyk}{0,0.85,0.87,0.35}
\definecolor{cafe}{cmyk}{0,0.72,1,0.45}
\definecolor{sepia}{cmyk}{0,0.83,1,0.70}
\definecolor{medianoche}{cmyk}{0.98,0.13,0,0.43}
\definecolor{grisclaro}{cmyk}{0,0,0,0.3}
\definecolor{grisoscuro}{cmyk}{0,0,0,0.5}
\definecolor{gris1}{gray}{0.1}
\definecolor{gris2}{gray}{0.2}
\definecolor{gris3}{gray}{0.3}
\definecolor{gris4}{gray}{0.4}
\definecolor{gris5}{gray}{0.5}
\definecolor{gris6}{gray}{0.6}
\definecolor{gris7}{gray}{0.7}
\definecolor{gris8}{gray}{0.8}
\definecolor{gris9}{gray}{0.9}

\definecolor{grisoscuro}{gray}{0.8}
\definecolor{grismedio}{gray}{0.7}
\definecolor{grisclaro}{gray}{0.5}
\definecolor{gris95}{gray}{0.95}
\definecolor{gris05}{gray}{0.05}

\usepackage{feyn}
\usepackage{amsmath,amssymb,graphics}
\usepackage[dvips]{graphicx}

\begin{document}

\begin{center}
{\Large \textbf{Higgs boson decay in the large $N$ limit}} \vskip1.0 cm {%
Rodolfo A. Diaz$^{1}$, Rafael Hurtado$^{1,2}$, R. Martinez$^{1}$,\ John
Morales$^{1,2}$ \\[0pt]
\textit{Universidad Nacional de Colombia$^{1}$ \\[0pt]
Departamento de F\'{\i}sica \\[0pt]
Bogot\'{a}, Colombia \\[0pt]
and \\[0pt]
Centro Internacional de F\'{\i}sica$^{2}$\\[0pt]
Bogot\'{a}, Colombia}} \vskip1.0 cm

\textbf{Abstract}
\end{center}

The Equivalence Theorem is commonly used to calculate perturbatively
amplitudes involving gauge bosons at energy scales higher than gauge boson
masses. However, when the scalar sector is strongly interacting the theory
is non-perturbative. We show that the Equivalence Theorem holds in the large
$N$ limit at next-to-leading order by calculating the decay widths $h \to
W^{+}W^{-}$ and $h\to \pi^{+}\pi^{-}$. We also show, in the same scheme of
calculations, that unitarity is fulfilled for the process $h\to
\pi^{+}\pi^{-}$ .



\vskip 1.0 cm %FT/GCP/2/03

%\newpage

\section{Introduction}

The Standard Model (SM) of the electroweak interactions, based on the $%
SU(2)_L\otimes U(1)_Y$ gauge symmetry \cite{1}, is a successful theory and
agrees with most experimental results \cite{2}. However, the scalar sector
responsible for the symmetry breaking of the SM is not well known and it has
not been tested yet. This sector gives masses to the particles of the model,
fermions and gauge fields, when the scalar field has a non vanishing Vacuum
Expectation Value (VEV) after the symmetry breaking. In the scalar sector a
Higgs particle appears with a mass given by $m_h^2=2 \lambda v^2$, where $%
\lambda$ is the coupling constant of the self-interacting term and $v$ is
the VEV ($v\approx 246$ GeV). $m_h$ is an unknown parameter so far.

Nevertheless, the precision tests of the SM impose strong bounds to the
Higgs mass when the scalar sector is weakly-coupled. The results from LEP
Electroweak Working Group analysis yield $m_{h}=114_{-45}^{+69}$ GeV (68\%
CL) \cite{LEP}, and an upper limit of $m_{h}<260$ GeV with one-sided 95$\%$
CL \cite{LEP}. The direct search of the Higgs boson done at LEP gives a
lower limit of $m_{h}>114.4$ GeV \cite{LEP}. On the other hand, it is
possible to have different models beyond the SM with a heavy Higgs with a
mass lying in the TeV scale for a strongly interacting scalar sector.
However, for this scenario to be held, the new physics contributions must
cancel those ones introduced by the heavy Higgs particle at low energies
\cite{pesking}.

If the SM is an effective theory derived from a more fundamental one, then
there is an associated $\Lambda$ scale for the appearance of new physics.
The use of theoretical arguments, like unitarity \cite{unitaridad},
triviality \cite{trivialidad} and vacuum stability \cite{estabilidad}, may
allow to get constraints for these two parameters $(\Lambda, m_h)$ \cite%
{kolda}.

The upper limit for the Higgs mass can be obtained by triviality
considerations in the Higgs sector \cite{calla}. When the quartic coupling
constant $\lambda$ in the scalar sector of the Higgs potential is
renormalized introducing a cut-off $\Lambda$, the coupling goes to zero when
$\Lambda$ goes to infinity, implying that $m_h$ goes to zero. This is not
the case for the SM, because it needs a massive scalar particle at low
energies to explain experimental results, and then the SM can be considered
as an effective theory below a given energy scale. If we knew this scale we
could predict the Higgs mass. Further, if the SM had a Higgs with a mass
around $1$ TeV, then the scalar sector would be strongly interacting and the
underlying theory would become non-perturbative\cite{urdiales}.

The amplitude for a heavy Higgs decaying into two longitudinally polarized gauge
bosons reads \cite{kniehl}
\begin{equation}
\mathcal{A} (h\rightarrow ZZ,WW)\approx \lambda (m_{h})\left( 1+2.8\frac{\lambda (m_{h})}{%
16\pi ^{2}}+62.1\left( \frac{\lambda (m_{h})}{16\pi ^{2}}\right) ^{2}\right) .
\end{equation}%
By considering that in the perturbation expansion the $\lambda ^{2}$ term
must be smaller than the $\lambda $ term, it is found that $\lambda
(m_{h})\approx 7$ implying that $m_{h}\approx 1$ TeV. On the other hand,
using the scattering process $WW\rightarrow ZZ$ mediated by a Higgs
particle, which might be important in future collider experiments like LHC
and linear colliders, the cross section for energies $\sqrt{s}>\!>m_{h}$ at
two loops level is given by \cite{maher}
\begin{equation}
\sigma (s)=\frac{1}{8\pi s}\lambda (s)^{2}\left[ 1-42.65\frac{\lambda (s)}{%
16\pi ^{2}}+2477.9\left( \frac{\lambda (s)}{16\pi ^{2}}\right) ^{2}\right] .
\end{equation}%
This cross section is negative for some values of $\lambda $ which means
that the perturbative expansion breaks down. Considering that the $\lambda
^{2}$ term must be smaller than the $\lambda $ term, a necessary condition
to have a convergent series is $\lambda \approx 4$, in this case $%
m_{h}\approx 700$ GeV \cite{riesselmann}. The above scenarios correspond to
the limit between weakly-coupled and strongly-coupled scalar sectors.

In the Marciano and Willenbrock paper \cite{marciano} they calculated the
decays of a heavy Higgs boson up to $\mathcal{O}(g^2m_h^2/m_W^2)$ in
perturbation theory using the Equivalence Theorem (ET) \cite{TE}, from which
the amplitude with gauge bosons longitudinally polarized at energies $%
\mathcal{O}(q^2>\!>m_W^2)$ is equivalent to the same amplitude but changing
the corresponding longitudinal components by the would-be Goldstone bosons.
For Higgs masses of the order of $m_h \approx 1$ TeV and $m_h \approx 1.3$
TeV the radiative corrections for the decay $h \to W^+ W^-$ are $7.3\%$ and $%
12\%$, respectively. At this scale the scalar sector is strongly-coupled and
the theory is non-perturbative. It is obvious that the amplitude at
next-to-leading order breaks the perturbative expansion because all Feynman
rules are proportional to the Higgs mass. For strongly interacting models is
necessary to use a non-perturbative method to calculate radiative
corrections and get bounded amplitudes. While it has been shown that the ET
holds order by order in perturbation theory, it has not been confirmed that
it does in non-perturbative calculations.

Due to the importance of studying the Higgs dynamics in non-perturbative
regimes, a formalism was introduced in Ref. \cite{Dobado} which uses Chiral
Perturbation Theory($\chi PT$) \cite{gasser}. Amplitudes are obtained as a
power expansion in the energy, this implies that the conventional ET does
not hold anymore \cite{veltman1}. Thus, a new formalism is necessary to have
an effective theory \cite{Dobado1}.

The large $N$ limit is an alternative approach that predicts bounded
positive defined amplitudes, consistent with pion dispersion \cite{john},
and useful to study the symmetry breaking of the strongly interacting sector
\cite{N}. The scalar sector of the SM can be modelled by a Linear Sigma
Model $O(4)$ and then generalized to a model with $O(N+1)$ symmetry. This
method has been applied to study the Higgs boson at TeV energy scales \cite%
{urdiales, apliN}. We show that the large $N$ limit can predict amplitudes
that fulfill the ET and the unitarity condition at next-to-leading order for
the SM, with a strongly interacting scalar sector, by using the $h \to W^{+}
W^-$ and $h \to \pi^+\pi^-$ processes.

In section 2 we introduce the Gauged Linear Sigma Model $O(N+1)$. In section
3 we calculate the Higgs decay widths, $h\rightarrow W^{+}W^{-}$ and $%
h\rightarrow \pi ^{+}\pi ^{-}$, in the large $N$ limit and we show that the
ET holds at next-to-leading order. In section 4 we show that the amplitude $%
h\rightarrow \pi ^{+}\pi ^{-}$ satisfies unitarity in the large $N$ limit.
In section 5 we give our conclusions.

\section{The $O(N+1)$ Model}

\hspace{12pt}

It is well known that the Linear Sigma Model represents the symmetry
breaking $O(N+1)\rightarrow O(N)$ with $N$ would-be Goldstone bosons which
belong to the fundamental irreducible representation of the remaining
symmetry $O(N)$. For the purposes of this work the would-be Goldstone bosons
will be named like pions $\pi $. For a gauge invariant model under $%
SU(2)_{L}\otimes U(1)_{Y}$ local symmetry the large $N$ limit for the SM is
defined as
\begin{equation*}
\mathcal{L}_{g}=\mathcal{L}_{YM}+(D_{\mu }\Phi )^{\dag }(D^{\mu }\Phi
)-V(\Phi ^{2})
\end{equation*}%
with $\Phi ^{\dag }=(\pi _{1},\pi _{2},\cdots ,\pi _{N},\sigma )$ and $\Phi
^{2}=\Phi ^{\dag }\Phi $. As usual $\mathcal{L}_{YM}$ is the Yang-Mills
Lagrangian of the SM and the covariant derivative is defined as
\begin{equation*}
D_{\mu }\Phi =\partial _{\mu }\Phi -ig\vec{T}^{L}\cdot \vec{W}_{\mu }\Phi
+ig^{\prime }T_{3}^{R}B_{\mu }\Phi ,
\end{equation*}%
where $\vec{T}^{L}=-(i/2)\vec{M}^{L}$ are the generators of the $SU(2)_{L}$
gauge group and $T_{3}^{R}=-(i/2)M^{Y}$ is the generator of the $U(1)_{Y}$
gauge group. The $M$ matrices are given by \cite{matriz}
\begin{equation*}
M_{ij}^{ab}=-i(\delta _{i}^{a}\delta _{j}^{b}-\delta _{i}^{b}\delta _{j}^{a})
\end{equation*}%
which belong to an irreducible representation of the $O(N+1)$ Lie algebra
with $i,j=1,2,3$ and $a,b=1,2,\hdots,N+1$. The matrices which belong to the
adjoint representation of the $SU(2)_{L}$ Lie algebra are given by
\begin{eqnarray*}
M_{1}^{L} &=&\begin{pmatrix} 0 & 0 & 0 & \cdots & - \\ 0 & 0 & - & \cdots &
0 \\ 0 & + & 0 & \cdots & 0 \\ \hdotsfor[2.5]{5} \\ \hdotsfor[2.5]{5} \\
\hdotsfor[2.5]{5} \\ + & 0 & 0 & \cdots & 0 \end{pmatrix}M_{2}^{L}=%
\begin{pmatrix} 0 & 0 & + & \cdots & 0 \\ 0 & 0 & 0 & \cdots & - \\ - & 0 &
0 & \cdots & 0 \\ \hdotsfor[2.5]{5} \\ \hdotsfor[2.5]{5} \\
\hdotsfor[2.5]{5} \\ 0 & + & 0 & \cdots & 0 \end{pmatrix} \\
&&
\end{eqnarray*}%
\begin{eqnarray*}
M_{3}^{L} &=&\begin{pmatrix} 0 & + & 0 & \cdots & 0 \\ - & 0 & 0 & \cdots &
0 \\ 0 & 0 & 0 & \cdots & + \\ \hdotsfor[2.5]{5} \\ \hdotsfor[2.5]{5} \\
\hdotsfor[2.5]{5} \\ 0 & 0 & - & \cdots & 0 \end{pmatrix} \\
&&
\end{eqnarray*}%
and, the corresponding matrix for the $U(1)_{Y}$ Lie algebra reads
\begin{equation*}
M^{Y}=\begin{pmatrix} 0 & + & 0 & \cdots & 0 \\ - & 0 & 0 & \cdots & 0 \\ 0
& 0 & 0 & \cdots & - \\ \hdotsfor[2.5]{5} \\ \hdotsfor[2.5]{5} \\
\hdotsfor[2.5]{5} \\ 0 & 0 & + & \cdots & 0 \end{pmatrix}
\end{equation*}%
where dots represent zeros. In this form we have a global $O(N+1)$ symmetry
with a local $SU(2)_{L}\otimes U(1)_{Y}$ symmetry.

The Higgs potential, invariant under $O(N+1)$, can be written as
\begin{eqnarray}
V(\Phi^2)=-\mu^2 \Phi^2 + \frac{\lambda}{4}(\Phi^2)^2.
\end{eqnarray}
Aligning the vacuum state as $\langle \phi \rangle_0 \equiv (0, \hdots ,v)$,
with $\Phi^2 = v^2 = 2\mu^2/\lambda$, the global symmetry $O(N+1)$ is broken
to $O(N)$ and the local symmetry is broken as $SU(2)_L \otimes U(1)_Y \to
U(1)_Q$. By defining the Higgs field as $h=\sigma - v$, we find the
following expression
\begin{eqnarray}
\mathcal{L}_g &=& \mathcal{L}_{YM}+\frac{1}{2}(D_\mu \pi_a)^\dag (D^\mu
\pi_a) + \frac{1}{2}(D_\mu h)^\dag (D^\mu h)  \label{mu} \\
&-& \frac{1}{2}m_h^2 h^2 - \lambda (\pi^2+h^2)^2 - 4 \lambda v h (\pi^2+h^2).
\notag
\end{eqnarray}

The gauge boson masses are obtained from the kinetic term,
\begin{eqnarray}
\frac{1}{2}\left(\frac{gv}{2}\right)^2 W_{\mu}^a W_a^{\mu}+\frac{1}{2} \left(%
\frac{g^{\prime}v}{2}\right)^2B_{\mu}B^{\mu}-\frac{gg^{\prime}v^2}{4}%
W_{\mu}^3B^{\mu}
\end{eqnarray}
where the mass eigenstates are given by
\begin{eqnarray}
W^+_\mu&=&(W^1_\mu-iW^2_\mu)/\sqrt{2}  \notag \\
W^-_\mu&=&(W^1_\mu+iW^2_\mu)/\sqrt{2}  \label{fisi} \\
Z_{\mu}&=&\cos\theta_W W_{\mu}^3-\sin\theta_W B_{\mu}  \notag \\
A_{\mu}&=&\sin\theta_W W_{\mu}^3+\cos\theta_W B_{\mu}  \notag
\end{eqnarray}
and $\theta_W$ is the Weinberg angle with $\tan\theta_W=g^{\prime}/g$. The $%
W^{\pm}_\mu$ fields, with $m_W=gv/2\approx 80.6 $ GeV masses, are the
charged gauge bosons, and the $Z_{\mu}$ field, with $m_Z = v (g^2 + {%
g^{\prime}}^{2})^{1/2}/2 \thickapprox 91.2$ GeV mass, is the weak neutral
gauge boson. The $A_{\mu}$ field is the massless photon.

The Lagrangian has terms of the form $g^2v\partial^{\mu}\pi_aW_{\mu}^a/4$,
mixing gauge bosons with would-be Goldstone bosons, which can be cancelled
by gauge fixing. We choose the Landau gauge ($\xi=0$) because in this gauge
a lot of Feynman diagrams cancel or suppress, the $\pi_a$ fields do not
couple to the ghost fields, and their propagators are massless. The final
Lagrangian can be written as
\begin{eqnarray}
\mathcal{L}[\pi,\vec{W},B,h] &=& -\frac{1}{2}\pi_a \Box \pi_a-\frac{1}{2}%
h(\Box+m_h^2)h-\lambda(\pi_a^2+h^2)^2  \notag \\
&-& 4\lambda vh(\pi_a^2+h^2) -\frac{g}{2} \partial^{\mu}\pi_1(W_{\mu}^3%
\pi_2-W_{\mu}^2\pi_3)  \notag  \label{jn} \\
&-& \frac{g}{2}\partial^{\mu}\pi_2(W_{\mu}^1\pi_3-W_{\mu}^3\pi_1) -\frac{g}{2%
}\partial^{\mu}\pi_3 (W_{\mu}^2\pi_1-W_{\mu}^1\pi_2)  \notag \\
&+& g\partial^{\mu}h(\vec{W}_{\mu}\cdot\vec{\pi}) -\frac{g^{\prime}}{2}
(\pi_1\partial_{\mu}\pi_2-\pi_2\partial_{\mu}\pi_1)B^{\mu}
-g\partial_{\mu}h\pi_3B^{\mu}  \notag \\
&+& \frac{1}{2}m_W^2 \vec{W}_{\mu}\cdot \vec{W}^{\mu}+\frac{1}{2}%
m_B^2B_{\mu}B^{\mu} -m_Wm_BW_{\mu}^3B^{\mu}  \notag \\
&+& \frac{g^2}{8}(\vec{W}_{\mu}\cdot\vec{\pi})(\vec{W}^{\mu}\cdot\vec{\pi}) +%
\frac{{g^{\prime}}^2v}{4}hB_{\mu}B^{\mu}  \notag \\
&+& \frac{{g^{\prime}}^2}{8}H^2B_{\mu}B^{\mu} -\frac{gg^{\prime}}{4}%
h^2W_{\mu}^3B^{\mu}-\frac{gg^{\prime}v}{2}hW_{\mu}^3B^{\mu}  \notag \\
&+& \frac{g^2}{8}h^2\vec{W}_{\mu}\cdot\vec{W}^{\mu}+\frac{g^2v}{4}h\vec{W}%
_{\mu}\cdot\vec{W}^{\mu} +\frac{{g^{\prime}}^2}{8}B_{\mu}B^{\mu}\vec{\pi}%
\cdot\vec{\pi}  \notag \\
&+& \frac{gg^{\prime}}{4}W_{\mu}^3B^{\mu}\vec{\pi}\cdot\vec{\pi} -\frac{%
gg^{\prime}}{2}\pi_3B_{\mu}(W_1^{\mu}\pi_1+W_2^{\mu}\pi_2)  \notag \\
&+& g^{\prime}m_WB_{\mu}(W_1^{\mu}\pi_2-W_2^{\mu}\pi_1)+ \frac{gg^{\prime}}{2%
}B_{\mu}(W_1^{\mu}\pi_2-W_2^{\mu}\pi_1)h  \notag \\
&+& \mathcal{L}_{YM} .
\end{eqnarray}
Thus we have a gauge theory spontaneously broken with the $\vec{\pi}=(\pi_1,
\pi_2, \pi_3)$ fields as the would-be Goldstone bosons of the broken
symmetry $SU(2)_L \otimes U(1)_Y/U(1)_Q$ and $\pi_a$ fields as the would-be
Goldstone bosons of the broken global symmetry $O(N+1)/O(N)$.

The theory for the large $N$ limit makes sense when $N \to \infty$ and gives
rise to finite amplitudes for different processes. To get finite amplitudes
is necessary to choose appropriate parameters in the large $N$ limit. We
will take the following definition
\begin{equation}
\lambda \approx 1/N
\end{equation}
in order to use perturbative expansion of the strongly interacting sector as
a function of the $\lambda$ parameter. With this definition, physical masses
must be finite and independent of $N$ in the large $N$ limit. From the
masses
\begin{eqnarray}
m_h^2 &=& 2\lambda v^2 \approx const  \notag \\
m_W^2 &=& \frac{g^2v^2}{4}\approx const  \notag \\
m_Z^2 &=& \frac{(g^2+{g^{\prime}}^2)v^2}{4}\approx const
\end{eqnarray}
we obtain for the other parameters of the model in the large $N$ limit the
following values
\begin{equation}
v\approx \sqrt{N} \;\; , \;\; g\approx 1/\sqrt{N} \; , \;\;
g^{\prime}\approx 1/\sqrt{N} .
\end{equation}

Finally, we obtain the Feynman rules necessary to calculate the decay widths
for $h \to W^{+}W^{-}$ and $h \to \pi^{+}\pi^{-}$ in the large $N$ limit,
see Fig. 1. \vskip 0.5cm
\begin{eqnarray*}
\Diagram{\vertexlabel^{w_{\mu}^+} \\ gd \\ & h \vertexlabel^h \\
\vertexlabel_{w_{\mu}^-} gu \\} = \displaystyle ig_{\mu \nu} \frac{gm_W}{%
\sqrt{N}} \hskip 2.0cm \Diagram{\vertexlabel^{\pi_a} \\ fd \\ & h
\vertexlabel^h \\ \vertexlabel_{\pi_b} fu \\} = \displaystyle -i\frac{g m_h^2%
}{2 m_W \sqrt{N}}\delta_{ab}
\end{eqnarray*}
\vskip0.2cm
\begin{eqnarray*}
\Diagram{\vertexlabel^{\pi^+} \\ fd \\ & h \vertexlabel^h \\
\vertexlabel_{\pi^-} fu \\} = \displaystyle -i\frac{g m_h^2}{2 m_W \sqrt{N}} %
\hskip 2.0cm \Diagram{ \vertexlabel^{\pi_a} \\ fd
fu^{^{^{^{^{^{\vertexlabel^{\pi^+}}}}}}} \\ \vertexlabel_{\pi_b} fu fd
\vertexlabel_{\pi^-} \\} = \displaystyle -i\frac{g^2 m_h^2}{4 m_W^2 N}
\delta_{ab}
\end{eqnarray*}
\vskip0.2cm
\begin{eqnarray*}
\Diagram{\vertexlabel^{\pi_a} \\ fd fu^{^{^{^{^{^{\vertexlabel^{\pi_c}}}}}}}
\\ \vertexlabel_{\pi_b} fu fd \vertexlabel_{\pi_d} } = \displaystyle -i\frac{%
g^2 m_h^2}{4 m_W^2 N} \delta_{ab} \delta_{cd} \hskip 1.5cm \feyn{h
a\vertexlabel^p \vertexlabel_h h} &=& \displaystyle{\frac{i}{p^2-m_h^2}}
\end{eqnarray*}
\vskip0.2cm
\begin{eqnarray*}
\feyn{f a\vertexlabel^p \vertexlabel_{\pi^{\pm},\pi_a} f} &=& \displaystyle{%
\frac{i}{p^2}}
\end{eqnarray*}
\vskip 0.5cm

\begin{center}
{\footnotesize Figure 1. Feynman rules in the Landau gauge for the SM in the
large $N$ limit }
\end{center}

\section{The Higgs Boson Decay and the Equivalence Theorem}

The SM in the large $N$ limit is associated with the $O(N+1)/O(N)$ and $%
SU(2)_L\otimes U(1)_Y/U(1)_Q$ global and local symmetry breaking schemes
respectively. We can calculate the amplitudes for $h\to W^+ W^-$ and $h \to
\pi^+ \pi^-$ decays in order to show that the ET holds in the proposed
scenario.

Feynman diagrams at tree level in this approximation are of the order of $%
\mathcal{O}(g)$ or $\mathcal{O}(g^{\prime})$ and of the order of $\mathcal{O}%
(1/\sqrt{N})$ in the large $N$ limit . The decay widths at tree level for $h
\to W^+ W^-$ and $h \to \pi^+ \pi^-$ processes are given by
\begin{eqnarray}
\Gamma (h \to W^+ W^-) &=& \frac{g^2 m_h^3}{64 \pi m_W^2} \left[1-\frac{%
4m_W^2}{m_h^2}\right]^{1/2}\left[1-\frac{4m_W^2}{m_h^2} +\frac{12m_W^4}{m_h^4%
}\right] ,  \notag \\
\Gamma(h \to \pi^+ \pi^-) &=& \frac{g^2 m_h^3}{64 \pi m_W^2}.
\end{eqnarray}

To obtain the amplitudes at next-to-leading order is necessary to introduce
the radiative corrections. First we calculate the self-energy of the scalar
particle $h$, whose Feynman diagrams at next-to-leading order are shown in
figure 2. In this case, the self-energy at one loop level with $\pi _{a}$
fields into the loops is of the order of $1/N$ times $N$ where $N$ is the
number of degrees of freedom running into the loop. Therefore, radiative
corrections are of the order of one in the large $N$ limit. The same
analysis can be done for the self-energy diagram with $l$ loops. It has two
vertices with $h\pi \pi $ and $l-1$ vertices with four $\pi _{a}$ and is of
the order of $(1/\sqrt{N})^{2}\dot{(}1/N)^{l-1}$ times $N^{l}$ the number of
pion fields running into the $l$ loops. Consequently, the self-energy
diagram with $l$ loops is of the order of one in the large $N$ limit.
However, the one irreducible particle function (1IP) for self-energy diagram with $W_{\mu }^{\pm
},Z_{\mu }$ into the loop is of the order of $1/N$, which is negligible in
the large $N$ limit.
\begin{eqnarray*}
&&-i\prod_{h}(q^{2})\Diagram{\Diagram{fs0=fs0\vertexlabel^h
h^{\vertexlabel^{\pi_a}} c_{_{\vertexlabel_{\pi_b}}} h \vertexlabel^h
fs0+fs0 \vertexlabel^h h^{\vertexlabel^{\pi_a}}
c_{_{\vertexlabel_{\pi_b}}}^{\vertexlabel^{\pi_c}}
c_{_{\vertexlabel_{\pi_d}}} h \vertexlabel^h fs0+fs0}} \\
&& \\
&&\Diagram{\Diagram{\vertexlabel^h h^{\vertexlabel^{\pi_a}}
c_{_{\vertexlabel_{\pi_b}}}^{\vertexlabel^{\pi_c}}
c_{_{\vertexlabel_{\pi_d}}}^{\vertexlabel^{\pi_e}}
c_{_{\vertexlabel_{\pi_f}}} h \vertexlabel^h fs0+ \cdots \cdots +fs0}
\underbrace{\Diagram{\vertexlabel^h
h^{\vertexlabel^{^{^{W_{\mu}^+}}}}}\hskip0.9cm\Diagram{gl glu
}\hskip0.9cm\Diagram{\vertexlabel_{_{_{_{_{_{_{_{_{_{_{_{_{_{_{{W_{%
\mu}^-}}}}}}}}}}}}}}}} h^{\vertexlabel^{h}}fs0+ \cdots}}_{{\bf {\cal
O}(\frac{1}{N})}}}
\end{eqnarray*}

\begin{center}
{\footnotesize Figure 2. Next-to-leading order Feynman diagrams that
contribute to the self-energy of the Higgs boson in the large $N$ limit}
\end{center}

After doing all calculations by using dimensional regularization, with $%
d=4-\epsilon$ integrals from the loops, we find (see appendix)
\begin{eqnarray}
-i \Pi_h(q^2)=\frac{\frac{g^2m_h^4}{8m_W^2}I_q}{1-\frac{ig^2m_h^2}{8m_W^2}I_q%
}
\end{eqnarray}
with $I_q$ given by
\begin{eqnarray}
I_q = \frac{i}{16 \pi^2} \left( \Delta + 2 -\log \frac{q^2}{\mu^2}-i\pi
\right)
\end{eqnarray}
where $\Delta = 2/\epsilon+\log 4\pi-\gamma_\epsilon$ and $\mu$ is
the renormalization scale. The choice of the renormalization
scale $\mu$ is arbitrary. Therefore, we shall adopt $\mu \approx
1$ $TeV$ as a reasonable choice. We have taken into account that
only the one irreducible particle  functions are important in
perturbation theory for the renormalization of parameters such as
the mass and the wave function \cite{Peskin}.

From the self-energy calculation the wave function renormalization of the
Higgs boson can be obtained as
\begin{eqnarray}
\left. Z_h=1+\frac{\frac{g^2}{256m_W^2 \pi^2 q^2}}{\left( 1+\frac{g^2m_h^2}{%
128 m_W^2 \pi^2} \left( \Delta + 2 -\log \frac{q^2}{\mu^2} -i \pi \right)
\right)^2} \right\vert_{q^2=m_{hR}^2} .  \label{zeta}
\end{eqnarray}
The contributions to $\pi^{\pm}$, $W^{\pm}$ and $Z_\mu$ self-energies are
proportional to $1/N$ and in the large $N$ limit we obtain that
\begin{equation}
Z_{\pi^{\pm}} = Z_{W_{\mu}} = Z_{Z_\mu} = 1 .
\end{equation}

To calculate the Higgs decays at this order, vertex corrections have to be
included as well, as shown in Fig. 3.
\begin{eqnarray*}
(a)\hskip0.2cm \Diagram{\Diagram{\vertexlabel^h h p
\Diagram{gu^{^{^{^{^{^{\vertexlabel^{W_{\mu}^{+}}}}}}}} \\ gd
\vertexlabel_{W_{\nu}^{-}}}fs0 = fs0} \Diagram{\vertexlabel^h h
&gu^{^{^{^{^{^{\vertexlabel^{W_{\mu}^{+}}}}}}}} \\ &gd
\vertexlabel_{W_{\nu}^{-}}}fs0+fs0 \Diagram{\vertexlabel^h
h_{\vertexlabel_{_{_{_{_{W_{\mu}^+}}}}}} \hskip0.9cm \Diagram{gl
glu}\hskip0.9cm
\Diagram{\vertexlabel^{^{^{^{^{^{^{^{W_{\nu}^-}}}}}}}}gu^{^{^{^{^{^{%
\vertexlabel^{W_{\mu}^{+}}}}}}}} \\ gd \vertexlabel_{W_{\nu}^{-}}}}fs0+fs0}
\cdots
\end{eqnarray*}
\vskip0.2cm
\begin{eqnarray*}
(b)\hskip0.2cm \Diagram{\vertexlabel^{h} h p
\Diagram{fu^{^{^{^{^{^{\vertexlabel^{\pi^{+}}}}}}}} \\ fd
\vertexlabel_{\pi^{-}}}fs0=fs0 \vertexlabel^{h} h^{\vertexlabel^{\pi_a}}
c_{\vertexlabel_{\pi_b}} \Diagram{fu^{^{^{^{^{^{\vertexlabel^{\pi^{+}}}}}}}}
\\ fd \vertexlabel_{\pi^{-}}}fs0 + fs0 \vertexlabel^{h}
h^{\vertexlabel^{\pi_a}} c_{\vertexlabel_{\pi_b}}^{\vertexlabel^{\pi_c}}
c_{\vertexlabel_{\pi_d}} \Diagram{fu^{^{^{^{^{^{\vertexlabel^{\pi^{+}}}}}}}}
\\ fd \vertexlabel_{\pi^{-}}} fs0+fs0 \cdots}
\end{eqnarray*}
\begin{eqnarray*}
\Diagram{fs0+fs0 \Diagram{\vertexlabel^h
h_{\vertexlabel_{_{_{_{_{W_{\mu}^+}}}}}}}} \hskip0.9cm \Diagram{gl glu}\hskip%
0.9cm \Diagram{\vertexlabel^{^{^{^{^{^{^{^{W_{\nu}^-}}}}}}}}fu^{^{^{^{^{%
\vertexlabel^{\pi^{+}}}}}}} \\ fd
\vertexlabel_{\pi^{-}}}+\cdots
\end{eqnarray*}
\begin{center}
\vskip -2.0cm \hskip 2.4cm
\begin{pspicture}(0,0)(8,7.5) %\grilla
\rput*(0.5,4){+} \rput*(1.5,4.2){\scriptsize$h$}
\rput*(3.5,4){\scriptsize$h$} \rput*(2.7,4.6){\scriptsize$\pi^+$}
\rput*(4,2.8){\scriptsize$\pi^+$}
\rput*(2.7,3.35){\scriptsize$\pi^-$}
\rput*(4,5.2){\scriptsize$\pi^-$}
\psline[linestyle=dashed](1,4)(2,4)
\psline[linewidth=1pt](2,4)(4,5) \psline[linewidth=1pt](2,4)(4,3)
\psline[linestyle=dashed](3.3,3.35)(3.3,4.65)
\end{pspicture}
\end{center}
\begin{center}
\vskip -3.0cm {\footnotesize Figure 3. Feynman diagrams in the
large $N$ limit which contribute to vertex interactions. $(a)$ $h
\to W^+W^-$, $(b)$ $h \to \pi^+ \pi^-$.}
\end{center}

The radiative corrections of $hW^+W^-$ vertex displayed in Fig. 3(a) are
suppressed since they are of the order of $1/N^2$ becoming negligible in our
approximation. The $hW^+W^-$ vertex can be written at this level as
\begin{eqnarray}
\Diagram{\vertexlabel^h h p
\Diagram{gu^{^{^{^{^{^{\vertexlabel^{W_{\mu}^{+}}}}}}}} \\ gd
\vertexlabel_{W_{\nu}^{-}}}fs0 = fs0} ig_{\mu \nu}\frac{gm_W}{ \sqrt{N}}.
\end{eqnarray}
For the $h\pi^+\pi^-$ vertex corrections shown in Fig. 3(b), the pions into
the loops give the most important contributions and can be written as

\begin{equation*}
\Diagram{\vertexlabel^{h} h p
\Diagram{fu^{^{^{^{^{^{\vertexlabel^{\pi^{+}}}}}}}} \\ fd
\vertexlabel_{\pi^{-}}}}=\frac{-ig}{2m_{W}\sqrt{N}}\left[ \frac{1}{\frac{1}{%
m_{h}^{2}}+\frac{g^{2}}{128m_{w}^{2}\pi ^{2}}\left( \Delta +2-\log \frac{%
q^{2}}{\mu ^{2}}-i\pi \right) }\right]
\label{vertex}
\end{equation*}%
where the $W^{\pm }$ contributions into the loops are suppressed by a $1/N$
factor with respect to the $\pi _{a}$ contributions. Similarly, the
contribution with a higgs running into the loop is also suppressed by a $1/N$ factor.

The wave function renormalization of the Higgs particle Eq.(\ref{zeta}) and
the vertex radiative correction Eq. (\ref{vertex}) diverge. To obtain finite
amplitudes the Higgs mass has to be renormalized \cite{urdiales}
\begin{eqnarray}
\frac{1}{m_{h_R}^2} \equiv \frac{1}{m_h^2}+\frac{g^2(\Delta+2)}{128\pi^2m_W^2%
}.
\end{eqnarray}
The real part of the $Z_h$ function Eq.(\ref{zeta}) is given by
\begin{eqnarray}
Z_h^{1/2}&=&1+\frac{g^2m_{h_R}^2}{16^2m_W^2\pi^2}\times \\
&&\frac{\left[1-\frac{2g^2m_{h_R}^2}{128m_W^2\pi^2}\log \frac{m_{h_R}^2}{\mu^2}%
+ \frac{g^4m_{h_R}^4}{8^2m_W^416^2\pi^4}\left(\log \left( \frac{m_{h_R}^2}{%
\mu^2}\right)^2-\pi^2\right)\right]}{ \left| 1- \frac{g^2m_{h_R}^2}{%
128m_W^2\pi^2} \left( \log \frac{m_{h_R}^2}{\mu^2} +i \pi \right)
\right|^4 }. \notag  \label{reze}
\end{eqnarray}
In the same way, the real part of the vertex correction Eq.(\ref{vertex})
can be expressed as
\begin{eqnarray}
\Diagram{\vertexlabel^{h} hs p
\Diagram{fu^{^{^{^{^{^{\vertexlabel^{\pi^{+}}}}}}}} \\ fd
\vertexlabel_{\pi^{-}}}}&&=\frac{-igm_{h_R}^2}{2m_W \sqrt{N}}
\times
\label{rever} \\
&&\left[1+ \frac{ \frac{g^2m_{h_R}^2}{128m_W^2\pi^2}\log \frac{m_{h_R}^2}{\mu^2%
}-\frac{g^4m_{h_R}^4}{8^2m_W^4 16^2\pi^4}\left(\log \left( \frac{m_{h_R}^2}{%
\mu^2}\right)^2-\pi^2\right)}{ \left| 1- \frac{g^2 m_{h_R}^2}{128 m_W^2 \pi^2}%
\left(\log \frac{m_{h_R}^2}{\mu^2}+i\pi\right) \right|^2 }\right]
\notag
\end{eqnarray}
Hence, the vertex corrections Eq.(\ref{rever}) multiplied by the factors of
the renormalized wave functions $Z_h^{1/2} Z_{\pi}$ give rise to the $h \to
\pi^+\pi^-$ amplitude at next-to-leading order and can be written at $%
\mathcal{O}(g^2m_h^2/m_W^2)$ as
\begin{eqnarray}
\mathcal{A}(h\to\pi^+\pi^-) &=& \frac{gm_{h_R}^2}{2m_W\sqrt{N}} \left[1+\frac{%
\frac{g^2m_{h_R}^2}{128m_W^2\pi^2}\log
\frac{m_{h_R}^2}{\mu^2}}{\left| 1-
\frac{g^2 m_{h_R}^2}{128 m_W^2 \pi^2}\left(\log \frac{m_{h_R}^2}{\mu^2}%
+i\pi\right) \right|^2} \right.  \notag \\
&+& \left. \frac{\frac{g^2 m_{h_R}^2}{16^2 m_W^2 \pi^2}}{\left| 1- \frac{%
g^2m_{h_R}^2}{128m_W^2\pi^2} \left( \log \frac{m_{h_R}^2}{\mu^2}
+i \pi \right)\right|^4} \right] .  \label{AHpp}
\end{eqnarray}

The same procedure is done for the $h \to W^+ W^-$ amplitude where the
vertex corrections are multiplied by the factors of the renormalized wave
functions $Z_h^{1/2} Z_{W}$, and can be written at $\mathcal{O}%
(g^2m_h^2/m_w^2)$ as
\begin{eqnarray}
\mathcal{A}(h\to W^+ W^-)=\frac{gm_{h_R}^2}{2m_W\sqrt{N}} \left[1+\frac{\frac{%
g^2 m_{h_R}^2}{16^2 m_W^2 \pi^2}}{\left| 1-
\frac{g^2m_{h_R}^2}{128m_W^2\pi^2} \left( \log
\frac{m_{h_R}^2}{\mu^2} +i \pi \right) \right|^4} \right] .
\label{AHWW}
\end{eqnarray}

In order to show that the ET holds for non-perturbative next-to-leading
order, we calculate Higgs decays into gauge bosons and pions in the large $N$
limit. We then compare the decay widths as obtained from the decay
amplitudes for $h \rightarrow \pi^+ \pi^-$ and $h \rightarrow W^+ W^-$ in Eqs. (\ref%
{AHpp}) and (\ref{AHWW}) respectively. Such decay widths are given by
\begin{eqnarray}
\Gamma (h \rightarrow \pi^+ \pi^-)&=& \frac{g^2 m_{h_R}^3}{64 \pi
m_W^2 N} \left[1+\frac{\frac{g^2m_{h_R}^2}{64\pi^2 m_W^2}\log
\frac{m_{h_R}^2}{\mu^2}}{\left| 1-
\frac{g^2 m_{h_R}^2}{128\pi^2 m_W^2}\left(\log \frac{m_{h_R}^2}{\mu^2}%
+i\pi\right) \right|^4} \right.  \notag \\
&+& \left. \frac{\frac{g^2 m_{h_R}^2}{128\pi^2 m_W^2}}{\left| 1- \frac{%
g^2m_{h_R}^2}{128 \pi^2 m_W^2} \left( \log \frac{m_{h_R}^2}{\mu^2}
+i \pi \right)\right|^8} \right] .
\end{eqnarray}
and
\begin{eqnarray}
\Gamma (h \rightarrow W^+ W^-)&=& \frac{g^2 m_{h_R}^3}{64 \pi
m_W^2 N} \sqrt{1-4\frac{m_W^2}{m_{h_R}^2}}
\left(1-4\frac{m_W^2}{m_{h_R}^2}+12\frac{m_W^4}{m_{h_R}^4}\right)  \notag \\
& & \left[1+\frac{\frac{%
g^2 m_{h_R}^2}{128 \pi^2 m_W^2}}{\left| 1- \frac{g^2 m_{h_R}^2}{64
\pi^2 m_W^2} \left( \log \frac{m_{h_R}^2}{\mu^2} +i \pi \right)
\right|^4} \right].
\end{eqnarray}

In Fig. 4 we display the ratio $%
\Gamma (h \rightarrow \pi ^{+}\pi ^{-})/\Gamma (h \rightarrow
W^{+}W^{-})$ as a function of the Higgs mass including
next-to-leading order corrections. From this figure it can be seen
that such quotient tends to one for large Higgs masses
($m_{h_R}\gtrsim 4.5\ $TeV), showing the validity of the ET at
high energies.

\begin{figure}
\begin{center}
\includegraphics[angle=0,width=10.0 cm]{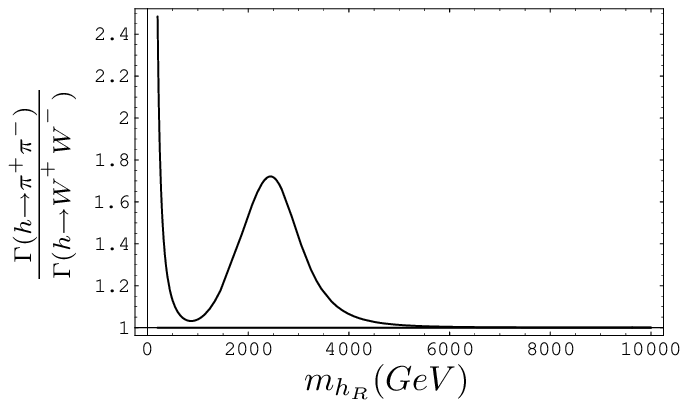}
\end{center}
\par
{\label{ejes}{\footnotesize Figure 4. The quotient $\Gamma
(h\rightarrow \pi ^{+}\pi ^{-})/\Gamma (h\rightarrow W^{+}W^{-})$
versus the renormalized Higgs mass, in the large $N$ limit at
next-to-leading order. This Fig. shows that both decay widths tend
to be equal for\ }}$m_{h_R}\gtrsim 4.5${\footnotesize \ TeV;
showing the validity of the ET at high energies.}
\end{figure}

\section{Unitarity in the large $N$ limit}

As a consequence of unitarity of the $S$-matrix, i. e. $S^{\dagger}S=1$, the
Optical Theorem is obtained . By defining $S=1+iT$, where the $T$ is called
the transition matrix, we have
\begin{eqnarray}
-i(T-T^{\dagger})=T^{\dagger}T  \label{T}
\end{eqnarray}
and since four momentum is conserved in the transition from initial state $%
\vert i \rangle$ to final state $\vert f\rangle$, we can always write
\begin{eqnarray}
\langle f \vert T \vert i \rangle = (2 \pi)^4 \delta^4(p_f-p_i)\mathcal{T}%
_{fi}
\end{eqnarray}
and
\begin{eqnarray}
\langle f \vert T^{\dagger} \vert i \rangle= \langle i \vert T \vert f
\rangle^{*} (2 \pi)^4 \delta^4(p_f-p_i)\mathcal{T}_{if}^{*}.
\end{eqnarray}
Inserting a complete set of intermediate states $\vert q \rangle$ we find
\begin{eqnarray}
\langle f \vert T^{\dagger} T \vert i \rangle = \sum_n \left(
\prod_{i=1}^{n} \int \frac{d^3 q_i}{(2\pi)^{3} 2E_i} \right)\langle f \vert
T^{\dagger} \vert q_i \rangle \langle q_i \vert T \vert i \rangle
\end{eqnarray}
and from the identity (\ref{T}) we can obtain the Cutkosky's rule\cite{cut}
\begin{eqnarray}
2 Im (\mathcal{T}_{if})=\sum_n \left( \prod_{i=1}^{n} \int \frac{d^3 q_i}{%
(2\pi)^{3} 2E_i}\right)\mathcal{T}_{f q_i}^* \mathcal{T}_{i q_i}(2
\pi)^4\delta^4(i \to \sum_{i}q_i)
\end{eqnarray}
where the sum runs over all possible sets of intermediate states $q_i$.

Applying this identity to the decay $\Gamma (h \to \pi^+ \pi^-)$ we find
\begin{eqnarray}
2 Im \left(\hskip0.1cm\Diagram{\vertexlabel^{p} hs p
\Diagram{fu^{^{^{^{^{^{\vertexlabel^{p^{+}}}}}}}} \\ fd
\vertexlabel_{p^{-}}}} \hskip0.1cm \right)&=& \int \frac{d^3q_a}{(2\pi)^3
2E_a}\frac{d^3q_b}{(2\pi)^3 2E_b}(2\pi)^4 \delta^4(p-q_a-q_b)  \notag \\
&\times& \left(\hskip0.1cm \Diagram{\vertexlabel^{p} hs p
\Diagram{fu^{^{^{^{^{^{\vertexlabel^{q_{b}}}}}}}} \\ fd
\vertexlabel_{q_{a}}}} \hskip0.1cm \right)^*\left(\hskip0.1cm %
\Diagram{\Diagram{\vertexlabel^{q_a}\\fd \\ \vertexlabel_{q_b}fu } p
\Diagram{fu^{^{^{^{^{^{\vertexlabel^{p^{+}}}}}}}} \\ fd
\vertexlabel_{p^{-}}}} \hskip0.1cm \right) .  \label{esta}
\end{eqnarray}
In the left-hand side of the previous equation we have the imaginary part of
the product of Eq. (\ref{vertex}) times the wave function $Z_h^{1/2} Z_{\pi}$%
, resulting
\begin{eqnarray}
2 Im (\mathcal{A}(h \to \pi^+ \pi^-))= \frac{\frac{g^3
m_{h_R}^4}{8m_W^3 16
\pi \sqrt{N}}} {\left|1-\frac{g^2m_{h_R}^2}{8m_W^2 16 \pi^2}\left( \log \frac{%
m_{h_R}^2}{\mu^2}+i \pi \right)\right|^2} .  \label{32}
\end{eqnarray}
For the right-hand side, we have to multiply the amplitudes calculated in
the large $N$ limit
\begin{eqnarray}
\Diagram{\vertexlabel^{h} h p
\Diagram{fu^{^{^{^{^{^{\vertexlabel^{q_{b}}}}}}}} \\ fd
\vertexlabel_{q_{a}}}fs0=fs0}Z_{h}^{1/2} Z_{\pi}\left( \Diagram{%
\vertexlabel^{h} h c \Diagram{fu^{^{^{^{^{^{\vertexlabel^{q_{b}}}}}}}} \\ fd
\vertexlabel_{q_{a}}}fs0+fs0}\Diagram{\vertexlabel^{h} h cc
\Diagram{fu^{^{^{^{^{^{\vertexlabel^{q_{b}}}}}}}} \\ fd
\vertexlabel_{q_{a}}}fs0+fs0 \cdots}\right)  \notag \\
\notag \\
\notag
\end{eqnarray}
i.e.,
\begin{eqnarray}
\mathcal{A}(h \to \pi_a \pi_b)&=&\frac{gm_{h_R}^2}{2m_W \sqrt{N}}\delta_{ab} %
\left[ \frac{1}{1-\frac{g^2m_{h_R}^2}{8m_W^2 16 \pi^2}\left( \log \frac{%
(q_a+q_b)^2}{\mu^2}+i \pi \right)} \right.  \notag \\
&+& \left. \frac{\frac{g^2m_{h_R}^2}{16^2 m_W^2 \pi^2 }}{\left(1-\frac{%
g^2m_{h_R}^2}{8m_W^2 16 \pi^2}\left( \log
\frac{m_{h_R}^2}{\mu^2}+i \pi \right)\right)^2}\right]
\end{eqnarray}
by
\begin{eqnarray}
\Diagram{\Diagram{\vertexlabel^{q_a}\\fd \\ \vertexlabel_{q_b}fu } p
\Diagram{fu^{^{^{^{^{^{\vertexlabel^{p^{+}}}}}}}} \\ fd
\vertexlabel_{p^{-}}}fs0=fs0}Z_{\pi}^2 \left(\hskip0.2cm \Diagram{\Diagram{%
\vertexlabel^{q_a}\\fd \\ \vertexlabel_{q_b}fu } c
\Diagram{fu^{^{^{^{^{^{\vertexlabel^{p^{+}}}}}}}} \\ fd
\vertexlabel_{p^{-}}}fs0+fs0} \Diagram{\Diagram{\vertexlabel^{q_a}\\fd \\
\vertexlabel_{q_b}fu } cc \Diagram{fu^{^{^{^{^{^{\vertexlabel^{p^{+}}}}}}}}
\\ fd \vertexlabel_{p^{-}}}fs0+fs0 \cdots} \right)  \notag \\
\notag \\
\notag
\end{eqnarray}
i.e.,
\begin{eqnarray}
\mathcal{A}(\pi_a \pi_b \to \pi^+ \pi^-)=\frac{g^2m_{h_R}^2}{4m_W^2 N}%
\delta_{ab} \left[ \frac{1}{1-\frac{g^2m_{h_R}^2}{8m_W^2 16
\pi^2}\left( \log \frac{(q_a+q_b)^2}{\mu^2}+i \pi \right)}\right]
.
\end{eqnarray}
Then the right-hand side of Eq. (\ref{esta}) becomes
\begin{eqnarray}
\left[\mathcal{A}(h \to \pi_a\pi_b) \right]^* \left[\mathcal{A}(\pi_a \pi_b
\to \pi^+ \pi^-) \right]&=& \frac{1}{(2\pi)^2}\times \frac{\frac{1}{4}\times\frac{g^3 m_{h_R}^4%
}{8m_W^3 \sqrt{N}}}{\left|1-\frac{g^2m_{h_R}^2}{8m_W^2 16
\pi^2}\left( \log
\frac{(q_a+q_b)^2}{\mu^2}+i \pi \right)\right|^2}  \notag \\
&=& \frac{1}{4} f(q_a,q_b)  \label{35}
\end{eqnarray}
where $1/4$ is the symmetry factor for identical bosons in the final state.

From Eq. (\ref{esta}) we define
\begin{eqnarray}
M&=&\frac{1}{4\pi^2}\int \frac{d^3q_a}{ 2E_a}\frac{d^3q_b}{2E_b}
\delta^4(p-q_a-q_b) \times  \notag \\
&&\left[\mathcal{A}(h \to \pi_a\pi_b) \right]^* \left[\mathcal{A}(\pi_a
\pi_b \to \pi^+ \pi^-) \right]  \label{36}
\end{eqnarray}
and the integral over $q_b$ can be written as
\begin{eqnarray}
\int \frac{d^3 q_b}{2 E_b}=\int_{-\infty}^{\infty}d^4 q_b \delta(q_a \cdot
q_a) \Theta(q_{b0}) .
\end{eqnarray}
Integrating the four-dimensional delta function in Eq. (\ref{36}) we obtain
\begin{eqnarray}
M&=&\frac{1}{16\pi^2}\int \frac{|q_a|^2 d|q_a| d \Omega}{2 E_a}\delta[%
(p-q_a)^2]\Theta(p_0-q_{a0})f(q_a,p-q_a)  \notag \\
&=& \frac{1}{16\pi^2}\int_0^E \frac{|q_a|dE_a d\Omega}{2}\delta[p^2-2p \cdot
q_a +q_a^2]f(q_a,p-q_a).  \label{38}
\end{eqnarray}
In the center-of-mass frame
\begin{eqnarray}
p=(E,\vec p), \hskip 0.5cm q_a=(E_a,\vec q_a)=(E^{\prime},\vec q), \hskip %
0.5cm q_b=(E_b,\vec q_b)=(E^{\prime},\vec q)
\end{eqnarray}
the integral in Eq.(\ref{38}) can be rewritten as
\begin{eqnarray}
M &=& \frac{1}{16\pi^2}\int_0^E \frac{|q_a|dE^{\prime}d\Omega}{2}\delta[%
E^2-2EE^{\prime}-p^2+2\vec p \cdot \vec q]f(q_a,p-q_a)  \notag \\
&=& \frac{g^3 m_{h_R}^4}{8 m_W^3 16 \pi^2
\sqrt{N}}\frac{|q_a|}{\left|1- \frac{g^2m_{h_R}^2}{8 m_W^2 16
\pi^2}\left(\log \frac{p^2}{\mu^2}+i\pi \right)\right|^2}\frac{d
\Omega}{2|-2E|}
\end{eqnarray}
and by using $p^2=m_{h_R}^2$
\begin{eqnarray}
M=\frac{\frac{g^3m_{h_R}^4}{8 m_W^3 16 \pi \sqrt{N}}}{\left|1- \frac{%
g^2m_{h_R}^2}{8 m_W^2 16 \pi^2}\left(\log
\frac{m_{h_R}^2}{\mu^2}+i\pi \right)\right|^2}.  \label{41}
\end{eqnarray}
Comparing equations (\ref{32}) and (\ref{41}) we see that the Higgs decay $%
\Gamma(h\to\pi^+\pi^-)$ calculated in the large $N$ limit at next-to-leading
order fulfills the unitarity condition.

\section{Conclusions}

We have shown that non-perturbative calculations at next-to-leading order in
the large $N$ limit for the case of a Higgs decaying into $W^{\pm }$ and $%
\pi ^{\pm }$ fulfill the ET. In particular, we found that the decay widths $%
\Gamma (h\rightarrow W^{+}W^{-})$ and$\ \Gamma (h\rightarrow \pi ^{+}\pi
^{-})$ get values that are basically identical for heavy Higgs bosons i.e. $%
m_{h_R}\gtrsim 4.5$ TeV.

On the other hand, we have also shown that calculations in the same scheme
for the Higgs decaying into pions respect unitarity. This results open the
possibility to study strongly interacting systems as could be the case of
the SM with a heavy Higgs boson.

\section*{Acknowledgments}

We thank to Alexis Rodriguez and Marek Nowakowski for reading the manuscript
and their comments. We also thank to COLCIENCIAS and DIB for their financial
support.

\section{Appendix}

In this appendix we show the explicit calculation of a Feynman diagram with $%
l$ loops in the large $N$ limit that contributes to the Higgs boson
self-energy.
\begin{eqnarray}
\Diagram{\vertexlabel^{h} hs} \underbrace{\Diagram{c c c \cdots c}}_{l-\text{%
loop%
%TCIMACRO{\U{b4}}%
%BeginExpansion
\'{}%
%EndExpansion
s}} \Diagram{hs \vertexlabel^{h}}&=&\frac{1}{2l}\left[\left(\frac{-igm_h^2}{%
2m_W\sqrt{N}}\right)(-N I_q)\right]\times\left[\left(\frac{-ig^2m_h^2}{%
4m_W^2 N}\right)\right]^{l-1}  \notag \\
&\times& (-N I_q)^{l-2}  \notag \\
&=&\frac{1}{2l}(-im_h^2)^{l+1}\left(\frac{-g^2 I_q}{4m_W^2}\right)^l
\end{eqnarray}
where $1/2l$ is the symmetry factor of the diagram. The first factor
corresponds to the initial and final loops times the vertices with three
particles, the second factor represents the product of the $l-1$ internal
vertices with four interacting fields and the last factor correspond to $l-2$
loops. Each loop contributes with an $N$ factor, as they have $N$
circulating pions.


\thebibliography{References}

\bibitem{1} S. Weinberg, {\em Phys. Lett.} {\bf 19}, 1264 (1967);
S.L. Glashow, {\em Nucl. Phys.} {\bf B20}, 579 (1961).

\bibitem{2} Particle Data Group, Phys. Rev. {\bf D66}, 1 (2002).

\bibitem{LEP} LEP Electroweak Working Group (2004),
http://lepewwg.web.cern.ch/LEPEWWG/.


\bibitem{pesking} R. S. Chivukula and N. Evans, Phys. Lett. {\bf B464}, 244 (1999);
R. S. Chivukula, N. Evans and C. Hoelbling. Phys. Rev. Lett. {\bf 85}, 511 (2000);
J. Bagger, A. Falk and M. Schwartz, Phys. Rev. Lett. {\bf 84}, 1385 (2000);
Michael E. Peskin, Phys. Rev. {\bf D64}, 093003 (2001).

\bibitem{unitaridad} B. Lee, C. Quigg and H. Thacker, Phys. Rev. {\bf D16}, 1519 (1977);
W. Marciano, G. Valencia and S. Willenbrock, Phys. Rev. {\bf D40}, 1725 (1989).

\bibitem{trivialidad} N. Cabibbo, L. Maiani, G. Parisi and R. Petronzio, Nucl Phys {\bf B158}, 295 (1979).

\bibitem{estabilidad} M. Sher, Phys. Rep. {\bf 179}, 273 (1989);
M. Quiros, Perspectives on Higgs Physics II, Ed. G.L. Kane, World Scientific, Singapore [arXiv: hep-ph/9703412].

\bibitem{kolda} Christopher Kolda, Hitoshi Murayama, JHEP 0007 (2000) 035.


\bibitem{calla} D.J.E Callaway, Phys. Rep. {\bf 167}, 241 (1988);
T. Hambye and K. Riesselmann, Phys. Rev. {\bf D55}, 7255 (1997);
J.S. Lee and J.K. Kim, Phys. Rev. {\bf D53},  6689 (1996).

\bibitem{urdiales} A. Dobado, J. Morales, J.R. Pelaez, M.T.
Urdiales, Phys. Lett. {\bf B387},  563 (1996);
A. Ghinculov, T. Binoth, J.J. van der Bij, Phys. Rev. {\bf D57}, 1487 (1998).


\bibitem{kniehl} A. Ghinculov, Nucl. Phys. {\bf B455},  21 (1995);
A. Frink {\it et. al.}, Phys. Rev. {\bf D54} (1996) 4548-4560

\bibitem{maher} P. Maher, L. Durand and K. Riesselmann, Phys. Rev. {\bf D48}, 1061 (1993); {\bf 52}, 553 (1995);
K. Riesselmann, Phys. Rev. {\bf D53},  6626 (1996).

\bibitem{riesselmann} U. Nierste and K. Riesselmann, Phys. Rev. {\bf D53}, 6638 (1996).

\bibitem{marciano} W.J.Marciano and S.S.D. Willenbrock Phys. Rev. {\bf D37}, 2509 (1988).

\bibitem{TE} B.W.Lee, C.Quigg and H. Thacker Phys. Rev. {\bf D16},  1519 (1977).
J.M.Cornwall, D.N. Levin and G. Tiktopoulus, Phys. Rev. {\bf D10},  1145 (1974).
M.S. Chanowitz and M.K. Gallaird Nucl. Phys. {\bf B261}, 379 (1985).
G.K. Gounaris, R. Kogerler and H. Neufeld, Phys. Rev. {\bf D34},  3257 (1986).
%J. Bagger and C.Schmidt Phys. Rev. {\bf D41},(1990)264. \\
%H.J.He, Y.P. Kuang and X. Li Phys. Rev. lett. {\bf 69}, (1992) 2619; Phys. Rev. {\bf D49}, (1994) 4842. \\
H.J.He and W. Kilgore, Phys. Rev. {\bf D55}, 1515 (1997).

\bibitem{Dobado} A. Dobado and M. Herrero, Phys. Lett. {\bf B228}, 495 (1989).
J. Donoghue and C. Ramirez, Phys. Lett. {\bf B234}, 361 (1990).

\bibitem{gasser} J. Gasser and H. Leutwyler, Ann. of Phys. {\bf 158},  142 (1984).
S. Weinberg Phys, Rev. Lett. {\bf 19}, 1264 (1967);
M.S. Chanowitz, M. Golden and H. Georgi, Phys. Rev. {\bf D36}, 1490 (1987);
A. Longhitano, Phys. Rev. {\bf D22}, (1980) 1166; Nucl. Phys. {\bf B188},  118 (1981).

%\bibitem{ramirez}J. Donoghue and C. Ramirez, Phys. Rev. {\bf B234}, (1990) 361;\\
%A. Dobado and M. Herrero, Phys. Lett. {\bf B228}, (1989) 495; \\
%T. Appelquist and C. Bernard, Phys. Rev. {\bf D22}, (1980) 200.

\bibitem{veltman1} H. Veltman, Phys. Rev. {\bf D41}, 264 (1990).

\bibitem{Dobado1} A. Dobado, J.R. Pelaez and M. T. Urdiales, Phys. Rev. {\bf D56}, 7133 (1997).

\bibitem{john} A. Dobado, A. Lopez and J. Morales, Il Nuovo Cimento {\bf 108A}, 335 (1995);
A. Dobado and J. Morales, Phys. Rev. {\bf D52}, (1995) 2878.

\bibitem{N} S. Coleman, R. Jackiw and H. D. Politzer, Phys. Rev. {\bf D10} 2491 (1974);
%S. Coleman, Aspects of Simmetry, Cambridge University Press (1985); \\
Dolan and Jackiw, Phys. Rev. {\bf D9}, 3320 (1974);
%R.G. Root, Phys. Rev. {\bf D10}, (1974) 3322; Phys. Rev. {\bf D11}, (1975) 831; Phys. Rev. {\bf D12}, (1975) 448; \\
M. Kobayashi and T. Kugo, Prog. Theor. Phys. {\bf 54}, 1537 (1975);
W. Bardeen and M. Moshe, Phys. Rev. {\bf D28}, 1372 (1983).

\bibitem{apliN} R.Casalbuoni, D. Dominici and R. Gatto, Phys. Lett. {\bf B147}, 419 (1984);
M.B. Einhorn, Nucl. Phys. {\bf B246}, 75 (1984);
A. Dobado and J.R. Pelaez, Phys. Lett. {\bf B286}, 136 (1992).

\bibitem{matriz} H. Georgi, Lie Algebras in Particle Physics. Addison Wesley, {\it Frontiers in Physics}. New York 1982.

\bibitem{Peskin} M. E. Peskin, D. V. Schroeder, \emph{An introduction to quantum field theory}, Westview Press (1995), Chap. 7.


\bibitem{cut} R. E. Cutkosky, J. Math. Phys. {\bf 1}, 429 (1960).
\end{document}